\documentclass[11pt]{article}

\usepackage[T1]{fontenc}
\usepackage[utf8]{inputenc}
\usepackage{amsmath, amssymb}
\usepackage{geometry}
\usepackage{hyperref}
\usepackage{enumitem}
\usepackage{graphicx}
\usepackage{booktabs}
\usepackage{array}
\usepackage[numbers,sort&compress]{natbib}

\hypersetup{
  colorlinks=true,
  linkcolor=blue,
  citecolor=blue,
  urlcolor=blue
}

\graphicspath{
  {figures/}%
  {../Figures/}%
  {/Users/bozprog/Library/Mobile Documents/com~apple~CloudDocs/FIGURES/}%
}

\title{\textbf{Link Wars: The Semantic Crisis}\\[4pt]
\large Is the debate over or is it just beginning?}
\author{Paul Borrill \\
D{\AE}D{\AE}LUS / Open Compute Project \\
\texttt{paul@daedaelus.com}}
\date{March 2026}

\begin{document}

\maketitle

\begin{abstract}
For fifty years, networking has fragmented whenever new workloads exposed hidden assumptions about time, ordering, failure, and trust. Today the proliferation of proprietary scale-up fabrics, consortium links, and Ethernet variants reflects not merely performance optimization but an avoidance of hard semantic commitments: what exactly completes, what is atomic, what is observable, and what happens under partial failure. This paper argues that the current interconnect landscape---NVLink, UALink, Ultra Ethernet, AELink/{\AE}thernet, TTPoE, and classical RDMA---suffers from a \emph{semantic crisis}: vendor-specific divergence disguised as optimization. We trace this crisis to a single root cause, the Forward-In-Time-Only (FITO) category mistake embedded in every major fabric stack. We examine the evidence---aspirational completion guarantees in RDMA/libfabric, fire-and-forget semantics in GPU fabrics, opaque proprietary stacks, incompatible multi-cloud ordering models, and the hidden cost of universal fencing---and show how each pathology arises from the same failure to define explicit, testable link semantics from APIs to bits on the wire. We present the conjecture that RDMA achieves reliability through universal fencing that collapses concurrency into serialized checkpoints, and that precise, minimal link semantics can maintain transactional correctness without global barriers, much as superscalar architectures separated execution order from retirement guarantees. We describe how Open Atomic Ethernet (OAE) under the Open Compute Project addresses the semantic crisis through bilateral transaction primitives with explicit ordering, completion, and failure visibility. Drawing on Pat Helland's analysis of scalable OLTP isolation semantics (the ``BIG DEAL''), we show that the semantic crisis is not confined to interconnects but pervades the entire stack: the same FITO assumption that produces universal fencing at the link layer produces relaxed isolation at the database layer and idempotence requirements at the application layer. Finally, we assess whether convergence on a single open link standard is still possible or whether fragmentation is now structural.
\end{abstract}

\section{Introduction}
\label{sec:intro}

At the Chiplet Summit in Santa Clara on February~17, 2026, a panel titled ``Link Wars: The Semantic Crisis'' brought together representatives from Intel, Lightmatter, Vaire Computing, Algo-Logic, Meta/ESUN, and D{\AE}D{\AE}LUS/OCP to confront a question that the interconnect industry has been avoiding: \emph{Is the industry chasing multiple link standards rather than a common Ethernet API because the problem truly demands it---or because we have failed to define a link semantics API rigorously enough to support a single, open standard across clouds, AI/ML, edge, automotive, and defense?}

The question is not new. Every decade since Ethernet's invention in 1973 has produced at least one ``link war''---Token Ring versus Ethernet, ATM versus Fast Ethernet, InfiniBand versus 10\,Gb Ethernet, FibreChannel versus iSCSI. In each case, the eventual resolution was not that one technology was objectively superior, but that the industry consolidated around whichever standard made the fewest semantic commitments while delivering adequate bandwidth. The uncommitted semantics were then patched, layered, or silently ignored by higher-level software.

Today's fragmentation is different in a critical respect. The workloads driving it---large-scale AI/ML training, disaggregated memory, multi-cloud orchestration, safety-critical automotive and defense systems---do not merely need more bandwidth. They need \emph{explicit semantic guarantees} about completion, ordering, atomicity, and failure visibility. These guarantees are precisely what the current generation of link standards fails to provide.

This paper argues that the crisis is not primarily about bandwidth, latency, or even governance. It is about \emph{semantics}---specifically, about the industry's systematic avoidance of hard semantic commitments at the link layer. We trace this avoidance to a single, largely unexamined design choice that pervades every major fabric stack: the Forward-In-Time-Only (FITO) assumption, which we identify as a category mistake~\cite{lamport1978time, FischerLynchPaterson1985}. We show how the FITO category mistake produces the specific pathologies visible in today's interconnect landscape, and we describe an alternative approach---Open Atomic Ethernet (OAE)---that resolves the crisis by defining explicit, testable link semantics from software APIs down to bits on the wire.

\section{The Interconnect Landscape}
\label{sec:landscape}

To understand the semantic crisis, we must first survey the current contestants. Each represents a different point in the design space, but all share a common deficiency: none provides a complete, testable specification of what ``completion'' means at the link layer.

\subsection{NVLink}

NVIDIA's NVLink is a proprietary, GPU-centric scale-up fabric designed to connect GPUs within a single server or across a small cluster via NVSwitch~\cite{nvidia2023nvlink}. NVLink provides very high bandwidth (up to 900\,GB/s per GPU in the Blackwell generation) and low latency within its designed operating envelope. Its semantic model is tightly coupled to CUDA's execution model: memory operations are ordered within a CUDA stream but may be arbitrarily reordered across streams unless explicit synchronization primitives are used. The key semantic limitation is that NVLink's ordering and completion guarantees are \emph{internal to the NVIDIA ecosystem}. There is no published, independently testable specification of NVLink's failure semantics, completion model, or behavior under partial failure. The link operates within a single trust domain---the assumption that all endpoints are NVIDIA GPUs running NVIDIA firmware---and this assumption is never stated as a formal precondition.

\subsection{UALink}

The Ultra Accelerator Link (UALink) consortium~\cite{ualink2025spec} was formed to provide an open alternative to NVLink for scale-up AI accelerator interconnects. UALink aims to standardize the physical and link layers for connecting accelerators within a pod. The consortium includes AMD, Broadcom, Cisco, Google, HPE, Intel, Meta, and Microsoft, representing a broad coalition of NVLink competitors. However, UALink's semantic commitments are still evolving. The published specifications focus on physical-layer interoperability and basic data transfer, but the crucial questions---what does completion mean? what ordering is guaranteed? what happens under link flap?---remain underspecified or deferred to higher layers.

\subsection{Ultra Ethernet (UEC)}

The Ultra Ethernet Consortium (UEC)~\cite{uec2024spec} targets scale-out AI/HPC workloads by extending Ethernet with features tailored for collective operations, congestion control, and reliability. UEC is building on standard Ethernet but adding transport-layer semantics for AI training patterns. The semantic challenge for UEC is that it inherits Ethernet's fundamental best-effort delivery model and attempts to retrofit reliability and ordering guarantees at the transport layer. This creates a gap: the link layer makes no ordering or completion commitments, and the transport layer must compensate with mechanisms whose failure modes are not always visible to applications.

\subsection{AELink / {\AE}thernet}

AELink, also known as {\AE}thernet, represents a deeper reinvention of Ethernet under the Open Compute Project (OCP). Unlike UEC, which adds transport-layer features on top of standard Ethernet, {\AE}thernet redesigns the link itself to provide bilateral, atomic transaction semantics~\cite{oae2024spec, OCP_OAE_Project}. The key design principle is \emph{full reversibility in the link}: every transaction either completes atomically (both sides know the outcome within bounded time) or rolls back cleanly. This is the approach developed by the Open Atomic Ethernet (OAE) project, which we examine in detail in Section~\ref{sec:oae}.

\subsection{TTPoE}

Tesla's Transport Protocol over Ethernet (TTPoE) is designed for Tesla's Dojo supercomputer. TTPoE replaces TCP's connection-oriented semantics with a simpler, lower-latency protocol optimized for the highly regular communication patterns of neural network training. Like NVLink, TTPoE operates within a single trust domain (Tesla's own datacenter) and makes semantic commitments that are optimized for a specific workload rather than general-purpose use.

\subsection{Classical RDMA}

Remote Direct Memory Access (RDMA), as specified by the InfiniBand Architecture~\cite{infiniband2024spec} and exposed through libfabric~\cite{ofiwg2024libfabric, grun2015brief}, is the most widely deployed high-performance fabric abstraction. RDMA's semantic model allows one-sided memory operations (read, write, atomic) that bypass the remote CPU. The completion model is based on completion queues (CQs) and work completions (WCs).

RDMA's semantic crisis is particularly instructive. The InfiniBand specification provides ordering guarantees through fences: a fence on a work request ensures that all prior operations in the same queue pair are visible before the fenced operation begins. In practice, RDMA implementations fence aggressively---often on every operation---to guarantee that any thread observing a completion is certain to observe every prior write~\cite{kalia2016design, mitchell2013using, dragojevic2014farm}. This universal fencing binds completion observability to strict prior-write visibility, collapsing concurrency into serialized checkpoints.

\subsection{The Common Deficiency}

Despite their differences in bandwidth, latency, topology, and governance model, all of these link technologies share a common deficiency: \emph{none provides a complete, testable, independently verifiable specification of its semantic guarantees}. NVLink's semantics are proprietary. UALink's are underspecified. UEC's are layered on top of a best-effort link. TTPoE's are workload-specific. RDMA's are buried in implementation-dependent fencing behavior. And even {\AE}thernet, while aspiring to explicit semantics, is still in development.

The question is: why? Why does every generation of interconnect technology repeat the same pattern of semantic underspecification?

\section{The Semantic Crisis}
\label{sec:crisis}

Modern interconnects suffer from vendor-specific semantic divergence disguised as optimization. The evidence is systematic and spans every major fabric family.

\subsection{RDMA and libfabric: Aspirational Completion}

The OpenFabrics Interfaces (OFI) framework, exposed through libfabric, provides a provider-agnostic API for high-performance fabric operations~\cite{ofiwg2024libfabric, grun2015brief}. The API specifies completion semantics, ordering flags, and memory registration requirements. However, these specifications are ``aspirational more than operational'': providers may silently degrade when hardware cannot deliver the requested semantics. A libfabric application requesting strict ordering may receive best-effort ordering on one provider and true strict ordering on another, with no programmatic way to detect the difference.

This is not a bug in any particular provider. It is a structural consequence of abstracting over semantically incompatible fabrics. Libfabric's design philosophy assumes that fabric providers can be treated as interchangeable beneath a common API. But semantic guarantees are not performance parameters that can be approximated---ordering is either guaranteed or it is not.

\subsection{GPU Fabrics: Fire-and-Forget Execution}

CUDA's execution model~\cite{nvidia2023nccl} relies on streams and events for ordering. Within a stream, operations execute in issue order. Across streams, there are no ordering guarantees unless explicit synchronization (events, stream synchronization) is used. The NVIDIA Collective Communications Library (NCCL) builds collective operations (all-reduce, all-gather, broadcast) on top of this model.

The semantic problem is that \emph{issue order is conflated with completion order}. When a CUDA kernel launches a series of NCCL operations, the programmer sees them issued in sequence. But completion---the point at which the results are visible to all participants---is not ordered by issue time. Fire-and-forget execution creates a gap between what the programmer \emph{believes} has happened and what has actually been committed. In normal operation, this gap is invisible because hardware timing coincidentally aligns issue order with completion order. Under stress, reordering, or partial failure, the gap becomes a source of silent corruption.

\subsection{Proprietary Stacks: Opaque Semantics}

NVLink and similar proprietary interconnects operate as closed systems. Their semantic guarantees are not published in a form that permits independent verification. This is not merely a business decision about intellectual property---it is a semantic deficiency. A link whose ordering and failure behavior cannot be reasoned about externally cannot be composed with other links in a heterogeneous system. The only safe assumption about an opaque link is that it provides no guarantees at all, which forces every higher layer to implement its own ordering, completion, and failure detection.

\subsection{Multi-Cloud: Semantic Babel}

In multi-cloud deployments, different cloud providers offer different completion, ordering, and failure semantics for what are nominally the same operations. An RDMA write on AWS (via Elastic Fabric Adapter and libfabric) has different completion semantics than the same operation on Azure (via InfiniBand) or GCP (via gVNIC). Applications that must span clouds cannot rely on any common semantic contract at the fabric layer. The result is that multi-cloud applications must implement their own consistency layer above the fabric, negating much of the performance advantage of RDMA in the first place.

\subsection{Universal Fencing: The Hidden Cost}

The most revealing symptom of the semantic crisis is RDMA's reliance on universal fencing. The InfiniBand specification defines fence semantics to ensure ordering: a fenced operation guarantees that all prior operations in the same queue pair are visible before the fenced operation begins~\cite{infiniband2024spec}. In principle, fences are a selective tool---a programmer should fence only when ordering is required.

In practice, RDMA systems fence aggressively, often on every operation~\cite{kalia2016design}. The reason is defensive: because the link's completion semantics are not fully explicit, the safest strategy is to ensure that every completion implies full prior-write visibility. This universal fencing binds completion observability to strict prior-write visibility, which collapses concurrency into serialized checkpoints. The resulting serialization overhead is significant but hidden, because it is embedded in the fabric stack rather than visible in application code.

\section{The FITO Category Mistake}
\label{sec:fito}

The semantic pathologies described in Section~\ref{sec:crisis} are not independent failures of different engineering teams. They are symptoms of a single, systemic design error: the Forward-In-Time-Only (FITO) assumption.

\subsection{What FITO Assumes}

Every major fabric protocol stack---from Shannon's 1948 channel model~\cite{lamport1978time} through TCP/IP, InfiniBand, libfabric, CUDA streams, and NVLink---assumes a communication model in which:

\begin{enumerate}[leftmargin=*]
\item A sender transmits a message.
\item The message propagates through a channel (which may lose, reorder, or corrupt it).
\item A receiver accepts the message (or doesn't).
\item The sender learns the outcome only by waiting for a subsequent, independent message from the receiver.
\end{enumerate}

This is the FITO model: information flows forward in time from sender to receiver, and the sender has no \emph{bilateral, bounded-time} knowledge of outcome. The sender's knowledge of completion depends on a separate acknowledgment message that is itself subject to the same channel uncertainties.

\subsection{Why FITO Is a Category Mistake}

FITO is typically treated as a physical constraint---a necessary consequence of finite signal propagation speed. But this is a category mistake: confusing a design choice with a physical law.

Relativistic causality constrains the \emph{speed} of information transfer. It does not constrain the \emph{structure} of a protocol boundary. There is no physical law that requires a link protocol to be unilateral (send, then hope for acknowledgment). Bilateral resolution---where both parties reach common knowledge of outcome at every round boundary, within bounded time---is physically realizable. It has been demonstrated at Layer~2 Ethernet in the Open Atomic Ethernet (OAE) project~\cite{oae2024spec, OCP_OAE_Project}.

The FITO assumption is invisible precisely because it is ubiquitous. It is embedded in Shannon's channel model, in Lamport's logical clocks~\cite{lamport1978time}, in the FLP impossibility proof~\cite{FischerLynchPaterson1985}, in TCP's three-way handshake, in RDMA's completion queue model, and in CUDA's stream semantics. We mistake the assumption for a constraint because we have never built a production link layer without it.

\subsection{How FITO Produces the Semantic Crisis}

Once the FITO assumption is made, every link protocol must compensate for the sender's uncertainty about completion. This compensation takes different forms in different stacks, but the pattern is universal:

\emph{RDMA compensates with fences.} Because the sender cannot know whether a write has been committed at the remote end without waiting for an independent completion, RDMA inserts fences to serialize operations and guarantee prior-write visibility. Universal fencing is the cost of FITO-based reliability.

\emph{GPU fabrics compensate with stream synchronization.} Because CUDA cannot guarantee that a remotely issued operation has completed without an explicit synchronization event, programmers must insert stream synchronization points. Fire-and-forget execution is what happens when this compensation is omitted.

\emph{Proprietary stacks compensate with closed trust domains.} Because FITO-based links cannot provide externally verifiable completion guarantees, vendors close the system: only endpoints running the vendor's firmware can participate, and the vendor implicitly guarantees semantics through homogeneity rather than through explicit specification.

\emph{Multi-cloud deployments have no compensation.} When applications span FITO-based fabrics from different vendors, there is no common semantic contract to compensate with. Each vendor's compensation mechanisms are vendor-specific, and the result is semantic Babel.

In every case, the root cause is the same: FITO forces the link to make \emph{weaker} semantic commitments than applications require, and the gap is filled by ad hoc, vendor-specific, often invisible compensation mechanisms.

\section{The Superscalar Conjecture}
\label{sec:conjecture}

We now state the central conjecture of this paper:

\begin{quote}
\emph{RDMA achieves reliability through universal fencing---binding completion observability to strict prior-write visibility. This collapses concurrency into serialized checkpoints. By instead defining precise, minimal link semantics, it is possible to maintain transactional correctness without imposing global barriers, increasing parallelism and reducing latency, much like superscalar architectures separated execution order from retirement guarantees.}
\end{quote}

The analogy to superscalar processors is deliberate and precise. In the 1980s, processor architects faced a similar semantic crisis. The von~Neumann model specified sequential execution: instruction $n$ completes before instruction $n{+}1$ begins. Early processors implemented this literally. Superscalar architectures broke this constraint by separating \emph{execution order} from \emph{retirement order}. Instructions could execute out of order, speculatively, and in parallel, provided that the retirement stage guaranteed that the programmer-visible state was consistent with sequential execution.

The key insight was that sequential execution was not a \emph{requirement} of correctness---it was a \emph{sufficient condition} for correctness that happened to be easy to implement. The actual requirement was much weaker: the programmer-visible state must be consistent with \emph{some} sequential execution. This weaker requirement could be satisfied while allowing massive internal parallelism.

We conjecture that the same decoupling is possible---and necessary---at the link layer. RDMA's universal fencing is analogous to in-order execution: it is a sufficient condition for correctness that collapses available concurrency. The actual requirement for transactional correctness at the link layer is much weaker than universal fencing. What is needed is:

\begin{enumerate}[leftmargin=*]
\item \textbf{Explicit ordering classes.} Not all operations require ordering. A link should support unordered, weakly ordered, and strictly ordered operations as distinct, application-selected contracts.
\item \textbf{Completion as a first-class semantic.} Completion must be explicitly defined, deterministically observable, and not conflated with buffer acceptance or issue acknowledgment.
\item \textbf{Failure visibility as a guarantee, not a hint.} Applications must be able to distinguish loss from reordering from failure, with bounded detection time.
\item \textbf{Minimal fencing.} Fences should be explicit, selective, and under application control---not universal and hidden in the fabric stack.
\end{enumerate}

These are precisely the semantic commitments that FITO-based links avoid, and precisely the commitments that OAE provides.

\section{Open Atomic Ethernet}
\label{sec:oae}

Open Atomic Ethernet (OAE) is a link-layer protocol under development within the Open Compute Project~\cite{OCP_OAE_Project, OCP_OAE_Wiki, oae2024spec}. OAE addresses the semantic crisis by replacing the FITO communication model with bilateral transaction primitives.

\subsection{Bilateral Transactions}

In OAE, every link-layer operation is a bilateral transaction: both endpoints participate in the operation and both reach a definite outcome (commit or abort) within bounded time. This is not a higher-layer protocol built on top of a FITO link---it is the fundamental operating mode of the link itself.

Bilateral transactions eliminate the completion ambiguity that pervades FITO-based links. When an OAE transaction commits, both sides \emph{know} it committed---not because they received an acknowledgment, but because the protocol's round structure guarantees common knowledge of outcome at every boundary. When a transaction aborts, both sides know it aborted, and the reason (link failure, contention, resource exhaustion) is explicitly communicated.

\subsection{Ordering Classes as Contracts}

OAE defines ordering as an explicit, application-selected contract rather than a fixed property of the link:

\emph{Unordered:} No placement or completion guarantees. Suitable for idempotent operations, telemetry, and best-effort traffic.

\emph{Weakly ordered:} Completions are ordered but payload delivery may reorder. Suitable for operations where completion order matters but data can arrive out of sequence.

\emph{Strictly ordered:} Both payload and completion are ordered. Suitable for operations that require full sequential consistency.

These ordering classes are not performance hints---they are testable contracts. An application that requests strict ordering receives strict ordering, and the link guarantees it. An application that needs only unordered delivery can operate without the overhead of fencing.

\subsection{Failure Visibility}

OAE provides deterministic failure visibility: applications can distinguish loss from reordering from failure, with bounded detection time. This is a strict improvement over RDMA's model, where failure detection depends on timeouts and the distinction between ``slow'' and ``failed'' is fundamentally ambiguous.

The key mechanism is that bilateral transactions have a bounded completion time. If a transaction does not complete within the bound, the link declares it failed and both sides are notified. There is no ambiguous state where one side believes a write has committed while the other side has not seen it.

\subsection{API-to-Wire Semantics}

The most distinctive feature of OAE's approach is that semantic guarantees are specified \emph{from the software API down to bits on the wire}. This is what we call ``API-to-wire'' semantics. The ordering, completion, and failure visibility guarantees that an application requests through the API are not interpreted by intermediate software layers---they are encoded in the link protocol itself and enforced by the hardware.

This stands in contrast to every other link technology surveyed in Section~\ref{sec:landscape}, where semantic guarantees are either unspecified at the link layer (Ethernet, UEC), specified but not verifiable (NVLink, UALink), or specified but routinely violated by provider-specific behavior (RDMA/libfabric).

\section{Scale-Up Versus Scale-Out: A False Dichotomy}
\label{sec:scaleup}

The interconnect industry currently treats scale-up (within a server or pod) and scale-out (across a datacenter or WAN) as fundamentally different problems requiring different link technologies. NVLink and UALink target scale-up; UEC and standard Ethernet target scale-out. We argue that this dichotomy is a symptom of the semantic crisis, not an inherent property of the problem.

Scale-up links enjoy two luxuries that scale-out links lack: short physical distances (which reduce latency and bit error rates) and homogeneous endpoints (which eliminate interoperability concerns). These luxuries allow scale-up links to make \emph{implicit} semantic guarantees---guarantees that hold because of physical properties of the deployment rather than explicit protocol specification.

But implicit guarantees are fragile. GPU-centric ordering assumes a single failure domain: all GPUs are either operational or the entire job fails. What happens when this assumption breaks---when one GPU in an NVLink mesh experiences a transient error, or when a pod must be drained for maintenance while training continues? NVLink's implicit semantics provide no answer, because the failure mode was never specified.

Scale-out links face the same semantic requirements as scale-up links---ordering, completion, atomicity, failure visibility---but cannot rely on implicit guarantees from physical proximity and endpoint homogeneity. The result is that scale-out deployments must implement these semantics in software, at enormous cost in latency, complexity, and failure opacity.

If link semantics were explicit and testable---if the link itself guaranteed ordering, completion, and failure visibility as specified contracts---then the distinction between scale-up and scale-out would reduce to a difference in physical parameters (latency, bandwidth, error rate) rather than a difference in semantic model. A single link standard with explicit semantics could serve both scale-up and scale-out, with physical parameters adapted to the deployment context.

\section{Governance and Fragmentation}
\label{sec:governance}

Why do new workloads always produce new fabrics instead of strengthening Ethernet semantics? The answer is partly economic (new consortia create new intellectual property) and partly political (incumbent vendors resist changes that commoditize their advantages). But there is a deeper, technical reason: \emph{Ethernet's semantic model is too weak to be strengthened incrementally}.

Standard Ethernet provides best-effort frame delivery. It guarantees nothing about ordering, completion, or failure visibility. Every attempt to add these guarantees---from TCP/IP through RDMA/RoCE to UEC---has been implemented as a \emph{layer on top of} Ethernet rather than a change to Ethernet itself. This layering preserves backward compatibility but prevents the link from making strong semantic commitments, because the link layer remains best-effort.

The result is a pattern that has repeated for decades: a new workload exposes the inadequacy of Ethernet's best-effort model; a new consortium forms to build a fabric that addresses the workload's specific needs; the new fabric makes proprietary or consortium-specific semantic commitments; and the industry fragments further.

OAE proposes to break this pattern by redesigning the Ethernet link layer itself. Rather than adding transport-layer patches to compensate for a weak link, OAE provides strong semantic guarantees at Layer~2. This is a more disruptive approach than UEC or UALink, but it addresses the root cause of fragmentation rather than adding another compensating layer.

Whether this approach can succeed politically---whether competing vendors can converge on observable, testable link semantics---is an open question. The technical argument is that explicit semantics \emph{reduce} the burden on vendors (because interoperability becomes testable rather than aspirational) and \emph{increase} the addressable market (because a single standard replaces multiple incompatible ones). But the political economy of standards bodies is not governed by technical arguments alone.

\section{Helland's Big Deal and the Database Evidence}
\label{sec:bigdeal}

The semantic crisis described in this paper is not confined to the link layer. The same structural pathology---weak semantic contracts compensated by ad hoc coping mechanisms---pervades the entire stack from interconnects through databases to applications. The most incisive analysis of this phenomenon at the database layer comes from Pat Helland's ``Scalable OLTP in the Cloud: What's the BIG DEAL?''~\cite{helland2024bigdeal}, presented at CIDR~2024 and subsequently discussed extensively in the D{\AE}D{\AE}LUS Mulligan Stew technical working group and the OCP Time Appliances Project.

\subsection{The BIG DEAL as Semantic Contract}

Helland's central observation is that scalable OLTP systems are split into databases and applications, and the thing that governs their interaction is not performance but \emph{isolation semantics}---what he calls ``the BIG DEAL.'' The BIG DEAL is the semantic contract between the database and the application: specifically, SQL with Read Committed Snapshot Isolation (RCSI). This contract defines what reads can see, what updates are atomic, and when completion is observable.

The parallel to interconnect semantics is exact. Just as the link layer defines the semantic contract between endpoints---what is ordered, what is complete, what is visible under failure---the database isolation level defines the semantic contract between storage and application. And just as link vendors avoid hard semantic commitments, database vendors have spent five decades \emph{relaxing} isolation semantics to achieve scale: from serializability to repeatable read to read committed to RCSI, each step trading semantic strength for concurrency.

\subsection{There Is No NOW}

Helland's most striking claim is that in a scalable BIG DEAL database, \emph{there is no NOW}~\cite{helland2024bigdeal}. There is no global ``current value'' of a record. Reads return snapshots---immutable views of the past, frozen at the reader's snapshot time. Updates create new record-versions visible to future snapshots. Time in the database is not a global coordinate but a partially ordered structure of commit times, snapshot times, and visibility boundaries.

This is the database-layer manifestation of the same insight that drives OAE at the link layer: global simultaneity is a fiction. In RDMA, the fiction is maintained by universal fencing---serializing all operations to create the illusion of a global ``now'' at which all writes are visible. In databases, the fiction was maintained by serializability until scale forced its abandonment. Helland's BIG DEAL makes the abandonment explicit: scalable databases have no NOW, only negotiated views of the past.

\subsection{FITO Coping Mechanisms at the Database Layer}

Helland's broader body of work---``Life Beyond Distributed Transactions''~\cite{helland2007life}, ``Memories, Guesses, and Apologies''~\cite{helland2009quicksand}, and ``Immutability Changes Everything''~\cite{helland2015immutability}---constitutes the most pragmatic and philosophically sophisticated attempt to build reliable systems within the FITO paradigm. His framework embraces uncertainty: local decisions are based on stale knowledge (memories), actions are taken without global truth (guesses), and compensating transactions clean up after failed predictions (apologies).

These are FITO coping mechanisms, precisely analogous to the link-layer coping mechanisms identified in Section~\ref{sec:crisis}:

\emph{Idempotence} is the application-layer analogue of RDMA fencing. Because a FITO sender cannot know whether a message was received, operations must be safe to retry. Idempotence makes forward retries harmless---but retries are only necessary because the communication model separates sending and receiving into two uncertain events rather than one bilateral transaction.

\emph{Immutability} is the storage-layer analogue of completion-queue serialization. Because FITO systems cannot retroactively correct the past, all state must be append-only. The log only grows; the arrow of time is baked into the data model. This prevents corruption but at the cost of ever-growing storage and the inability to reclaim resources without complex garbage collection.

\emph{Compensating transactions} (apologies) are the workflow-layer analogue of timeout-based failure detection. Because FITO systems discover errors only after the fact, they must undo incorrect actions through subsequent corrective actions rather than preventing the errors at the interaction boundary.

In every case, the pattern is identical: FITO forces the system to make weaker commitments than the application requires, and the gap is filled by increasingly baroque compensation mechanisms that add complexity, latency, and failure modes.

\subsection{Set Reconciliation and the Path Forward}

In a joint presentation to the OCP Time Appliances Project in December 2024, Helland and Daniel May explored practical rateless set reconciliation~\cite{helland2024tap}---an algorithm that can efficiently find differences between large sets in time proportional to the \emph{difference} rather than the set size. Their key insight was that set reconciliation, combined with time-bounded epochs, enables a fundamentally different approach to distributed agreement: rather than coordinating in advance (as consensus protocols require), nodes can reconcile after the fact with provably bounded cost.

This connects directly to the semantic crisis. The reason distributed systems need consensus, fencing, and serialization is that FITO communication provides no bilateral guarantee of what the other side knows. Set reconciliation offers a complementary approach: if both sides can efficiently determine what the other side \emph{doesn't} know, the need for preventive serialization diminishes. Combined with OAE's bilateral transaction primitives at the link layer, set reconciliation at the application layer suggests a path toward systems that achieve consistency without the overhead of universal fencing or global coordination.

\subsection{The Stack-Wide Pattern}

The Helland evidence confirms that the semantic crisis is not a link-layer accident but a stack-wide structural consequence of the FITO assumption:

At the \emph{link layer}, FITO produces universal fencing, fire-and-forget semantics, and opaque proprietary stacks (Section~\ref{sec:crisis}).

At the \emph{database layer}, FITO produces relaxed isolation, compensating transactions, and the abandonment of global consistency (the BIG DEAL).

At the \emph{application layer}, FITO produces idempotence requirements, retry storms, and the ``memories, guesses, and apologies'' framework.

At every layer, the root cause is the same: the communication primitive is too weak to support the semantic requirements of the layer above, and the gap is filled by ad hoc, layer-specific compensation. The implication is that fixing the link layer alone---while necessary---is not sufficient. But it is the right place to start, because every higher layer inherits the link's semantic weakness. A link that provides bilateral, atomic, semantically explicit transactions reduces the compensation burden at every layer above it.

\section{Is Convergence Possible?}
\label{sec:convergence}

We return to the question posed at the Chiplet Summit panel: is a single, broadly acceptable link standard still possible, or is fragmentation now structural?

The optimistic case rests on three observations. First, the semantic requirements are \emph{convergent}: every workload---AI/ML, cloud, edge, automotive, defense---needs some combination of ordering, completion, atomicity, and failure visibility. The specific mix varies, but the vocabulary is shared. Second, explicit semantics are \emph{composable}: a link that provides ordering classes as contracts can serve workloads with different ordering requirements simultaneously. Third, the cost of fragmentation is \emph{accelerating}: as systems become more heterogeneous and multi-vendor, the overhead of bridging semantically incompatible fabrics grows superlinearly.

The pessimistic case rests on history. Every previous link war was resolved not by semantic convergence but by market consolidation: one technology won enough market share that alternatives became economically unviable. The semantic gaps were never resolved---they were papered over by software layers that assumed the winning technology's specific behavior. There is no reason to expect the current generation to be different.

We believe the determining factor will be whether the industry recognizes the semantic crisis as a \emph{category mistake} rather than a market competition. If the debate remains framed as ``which link is fastest?,'' fragmentation will continue, because speed is a parameter that vendors optimize independently. If the debate shifts to ``which link provides testable semantic guarantees?,'' convergence becomes possible, because testability is a property that benefits from standardization.

The panel's most provocative question---``If bandwidth doubled tomorrow but semantic guarantees stayed ambiguous, would we actually have solved anything?''---points directly at this distinction. The answer, we contend, is no. The crisis is not about bandwidth or latency. It is about trust in completion. And trust cannot be doubled by doubling bandwidth.

\section{Conclusion}
\label{sec:conclusion}

The interconnect industry faces a semantic crisis that no amount of bandwidth will resolve. The proliferation of link standards---NVLink, UALink, UEC, AELink/{\AE}thernet, TTPoE, and classical RDMA---reflects not innovation but the industry's systematic avoidance of hard semantic commitments. Each new fabric compensates for the same underlying deficiency: the FITO assumption that forces links to make weaker guarantees than applications require.

The FITO category mistake---confusing a design choice with a physical constraint---has produced universal fencing in RDMA, fire-and-forget semantics in GPU fabrics, opaque proprietary stacks, and incompatible multi-cloud ordering models. These are not independent engineering failures. They are predictable consequences of building every link on a communication model that was never adequate for the semantic requirements of modern distributed systems.

Open Atomic Ethernet offers a path beyond the crisis by replacing the FITO model with bilateral transaction primitives that provide explicit, testable semantic guarantees from APIs to bits on the wire. Whether the industry will take this path depends on whether we can reframe the debate from ``which link is fastest?''\ to ``which link provides testable semantic guarantees?''

The debate, we submit, is just beginning.

\bigskip

\noindent\textbf{Acknowledgments.} The author thanks the Chiplet Summit 2026 panelists---Anjali Singhai Jain (Intel), Bijan Nowroozi (Lightmatter), Hannah Earley (Vaire Computing), John Lockwood (Algo-Logic), and Manoj Wadekar (Meta/ESUN)---and moderator Bill Wong (Electronic Design) for a discussion that sharpened several arguments in this paper. The OAE project is developed under the Open Compute Project. Related work appears in \emph{What Distributed Computing Got Wrong}~(arXiv:2602.18723), \emph{Circumventing FLP with OAE}~(arXiv:2602.20444), \emph{Lamport's Arrow of Time}~(arXiv:2602.21730), and \emph{Why Atomicity Matters to AI/ML}~(arXiv:2603.02603).

\appendix

\section{OAE Governance and Intellectual Property}
\label{app:governance}

The Open Atomic Ethernet (OAE) project operates under the Open Compute Project (OCP) with an explicit governance mission that distinguishes it from the consortium-driven approaches of UALink, UEC, and other industry groups.

\subsection{Open Governance and Return to IEEE Standards Stewardship}

A central objective of the OAE project is the return of Ethernet semantic standardization to IEEE stewardship. Ethernet originated as an IEEE standard (IEEE~802.3), and its physical and MAC layers remain under IEEE governance. However, the semantic extensions that modern workloads require---ordering, completion, atomicity, failure visibility---have proliferated outside IEEE, in vendor consortia (InfiniBand Trade Association, UALink Consortium, Ultra Ethernet Consortium) and proprietary stacks (NVLink, TTPoE). The OAE project's position is that semantic standardization belongs within IEEE, where it can benefit from IEEE's established processes for broad industry participation, technical review, and long-term stewardship.

The OCP serves as an incubation venue: the technical work is developed openly under OCP's collaborative framework, with the intent that mature specifications will be contributed to IEEE for formal standardization.

\subsection{Intellectual Property Licensing}

The intellectual property underlying the OAE specifications is intended---though not guaranteed---to be \emph{freely licensable} by implementations that conform to the resulting standards. Conforming implementations would receive access to the relevant IP at no cost, removing the licensing barriers that have historically fragmented fabric ecosystems and concentrated power in holders of essential patents.

For non-conforming implementations, the IP would be available under \emph{Reasonable and Non-Discriminatory (RAND)} terms, matching the licensing conventions long established by the IEEE Standards Association. RAND licensing ensures that even implementations that diverge from the standard can obtain access to the underlying IP on fair, published terms, while preserving the incentive for conformance.

This two-tier licensing model---free for conforming implementations, RAND for others---is designed to maximize adoption of the open standard while respecting the IP contributions of participants. It reflects the OAE project's conviction that open semantics create larger markets, not smaller ones, and that the industry's long-term interests are better served by broad, interoperable adoption than by proprietary fragmentation.

\section{The Atomic Ethernet Industry Association Manifesto}
\label{app:manifesto}

The following manifesto was published by the Atomic Ethernet Industry Association (AEIA) in January 2026~\cite{aeia2026manifesto}. It is reproduced here as a statement of principles that motivates and contextualizes the technical arguments of this paper.

\bigskip

\begin{center}
{\large\bfseries The Atomic Ethernet Manifesto}\\[4pt]
{\small D{\AE}D{\AE}LUS Team for the AEIA \quad January 2026}
\end{center}

\medskip

\noindent\textbf{Preamble.}
The computing industry is at an inflection point. As fabrics collapse from racks to boards, from boards to packages, and from packages to die-to-die chiplet domains, long-standing assumptions about networking, programmability, and system semantics no longer hold. Bandwidth alone is no longer the dominant constraint. Instead, semantic correctness, determinism, failure visibility, and programmability boundaries now define system scalability and trustworthiness.

Atomic Ethernet arises from a simple but foundational observation: distributed systems, databases, and accelerators are constrained not by what the network claims to do, but by what the bits on the wire actually guarantee.

This manifesto articulates why the industry must converge on a shared, open, and verifiable semantic contract between applications, SmartNICs, and Ethernet fabrics, and why that contract must be standardized across clouds, vendors, and deployment scales.

\medskip

\noindent\textbf{The Problem We Must Solve.}

\emph{The Semantic Gap.}
Modern software systems operate using rich semantic models, including transactions and atomicity, ordering and causality, completion and failure visibility, and memory consistency and isolation. Yet the underlying networks that support these systems expose incomplete, opaque, or vendor-specific behavior. The result is a widening gap between what applications assume and what the network can actually guarantee.

This gap is the root cause of systemic pathologies, including latent data corruption, heisenbugs in distributed systems, unrecoverable failures masked as best-effort behavior, and fragile retry logic that reintroduces inconsistency at scale.

\emph{The Mystery Stack.}
Many high-performance interconnects today present what can only be described as a mystery stack: proprietary stovepipes that cannot be reasoned about externally; opaque transport behavior hidden behind abstractions; divergent semantics across vendors for ostensibly similar operations; and application developers forced to reverse-engineer behavior through failure. This fragmentation suppresses innovation, limits portability, and concentrates power in closed ecosystems while increasing systemic risk.

\emph{Scaling Down Is Not Enough.}
Technologies designed for large, best-effort, multipath fabrics cannot simply be scaled down to chiplet-scale domains. As fabrics shrink, path diversity collapses, latency bounds tighten, synchronization becomes feasible, and failure modes change qualitatively. This shift is architectural, not incremental. It demands revisiting core assumptions rather than repackaging existing ones.

\medskip

\noindent\textbf{Our Core Belief.}
Atomic semantics must be native to the fabric. They cannot be reliably retrofitted above the network stack without reintroducing ambiguity, retries, and hidden state that violate the guarantees applications depend on. In particular: atomicity cannot be layered on top of unreliable substrates; ordering cannot be inferred from time alone; completion without semantics is meaningless; and reliability without observability is deception.

\medskip

\noindent\textbf{The Atomic Ethernet Vision.}

\emph{A Cross-Cloud Transaction Processing Interface.}
We call for an open, industry-standard Transaction Processing Interface (TPI) that explicitly binds application-visible semantics to SmartNIC behavior to observable wire-level guarantees. This interface defines a shared contract between software and hardware that is portable across vendors, clouds, and deployment scales.

\emph{Ordering Semantics.}
At a minimum, the interface must standardize the following ordering classes: \emph{Unordered} (no guarantees on payload placement or completion order); \emph{Weakly Ordered} (payload may be reordered, but completions are delivered in order); and \emph{Strictly Ordered} (payload placement and completion delivery are both ordered). These are semantic contracts, not performance hints.

\emph{Completion and Failure Models.}
Applications must be able to reason explicitly about completion and failure. This includes explicit versus implicit completion, fire-and-forget semantics with bounded failure signaling, tight latency bounds where feasible, and deterministic buffer lifecycle management. Completion without meaning is not completion.

\emph{Reliability as a First-Class Concept.}
Reliability must be designed, not assumed. This requires clear distinction between loss, reordering, and failure; mechanisms appropriate to single-path or tightly constrained fabrics; and explicit exposure of reliability guarantees to software.

\emph{Memory and Atomic Operations.}
The fabric must support well-defined memory semantics, one-sided operations with explicit visibility guarantees, and collectives and higher-level primitives built on verifiable foundations.

\medskip

\noindent\textbf{Why Openness Matters.}
An open semantic standard creates larger markets, not smaller ones. It enables true multi-cloud portability, allows independent innovation at the NIC, fabric, and application layers, and prevents semantic fragmentation disguised as optimization. History shows that closed semantics do not scale ecosystems.

\medskip

\noindent\textbf{Our Ambition.}
Atomic Ethernet aims to support Kubernetes-scale distributed systems, accelerator-class programming models, chiplet-scale fabrics, and future systems we cannot yet name. This is only possible if Layer~2 itself is trustworthy. Without atomic semantics at the fabric level, higher layers inherit ambiguity, debugging becomes archaeology, and correctness becomes probabilistic.

\medskip

\noindent\textbf{A Call to Action.}
We invite hardware architects, SmartNIC designers, cloud providers, database and distributed systems researchers, standards bodies, and open-source communities to participate in defining an industry-wide, open, and verifiable semantic foundation for Ethernet fabrics. This effort is not about faster links. It is about making the world safe for transactions.

\medskip

\noindent\textbf{Closing Statement.}
Atomic Ethernet is not a product. It is not a vendor position. It is not a single implementation. It is a commitment: the semantics programmers depend on must be as real, observable, and standardized as the bits on the wire. Anything less is technical debt we can no longer afford.

\bibliographystyle{plainnat}
\bibliography{references/link-wars-semantic-crisis}

\end{document}